\documentstyle[preprint,aps,psfig,amsmath]{revtex}

\begin{document}

\begin{center} Degeneracies when T=0 Two Body Interaction Matrix Elements 
are Set Equal to Zero : Talmi's method of calculating coefficients of fractional parentage to states forbidden by the Pauli principle.
\medskip \\ 
Shadow J.Q. Robinson$^{a}$ and Larry Zamick$^{a,b}$

\noindent a) Department of Physics and Astronomy,
Rutgers University, Piscataway, \\New Jersey  08855

\noindent b)  TRIUMF,4004 Wesbrook Mall, Vancouver, British 
Columbia, \\Canada, V6T 2A3

\end{center}

\bigskip
\begin{abstract}
In a previous work we studied the effects of setting all two body T=0
matrix elements to zero in shell model calculations for $^{43}$Ti
($^{43}$Sc) and $^{44}$Ti.  The results for $^{44}$Ti were
surprisingly good despite the severity of this approximation.  In this
approximation degeneracies arose in the T=$\frac{1}{2}$
I=$(\frac{1}{2})^-_1$ and $(\frac{13}{2})^-_1$ states in $^{43}$Sc and
the T=$\frac{1}{2}$ $I=(\frac{13}{2})_2^-$, $(\frac{17}{2})^-_1$, and
$(\frac{19}{2})_1^-$ in $^{43}$Sc.  The T=0 $3_2^+$, $7_2^+$, $9_1^+$,
and $10_1^+$ states in $^{44}$Ti were degenerate as well.  The
degeneracies can be explained by certain 6j symbols and 9j symbols
either vanishing or being equal as indeed they are.  Previously we
used Regge symmetries of 6j symbols to explain these degeneracies.  In
this work a simpler more physical method is used.  This is Talmi's
method of calculating coefficients of fractional parentage for
identical particles to states which are forbidden by the Pauli
principle.  This is done for both one particle cfp to handle 6j
symbols and two particle cfp to handle 9j symbols.  The states can be
classified by the dual quantum numbers ($J_{\pi},J_{\nu}$).

\end{abstract}
\vspace{2.5in}

\newpage

\section{Introduction}

In a previous work~\cite{dowreg} we considered both states in a single
j shell and a full FP calculation for T=$\frac{1}{2}$ states in
$^{43}$Ti ($^{43}$Sc) and T=0 states in $^{44}$Ti. We compared the
results in which the full interaction was used and one in which all
the T=0 two body interaction matrix elements were set equal to zero.
Despite the apparent severity of the latter, the results in $^{44}$Ti
were surprisingly good, especially for states of even angular
momentum.  The odd I states were somewhat too low and got shifted up
(almost uniformly) when the T=0 matrix elements were put back in.

When the T=0 matrix elements were set equal to zero certain
degeneracies appeared.  eg I=$(\frac{1}{2})^-_1$, $(\frac{13}{2})^-_1$
T=$\frac{1}{2}$ states in $^{43}$Ti($^{43}$Sc) as well as
T=$\frac{1}{2}$ I=$(\frac{13}{2})_2^-$, $(\frac{17}{2})^-_1$, and
$(\frac{19}{2})^-_1$. In $^{44}$Ti the 3$_2^+$,7$_2^+$, 9$_1^+$, and
10$_1^+$ states were degenerate.  These degeneracies required certain
6j and 9j symbols to vanish and others to be equal.  This was verified
by simply looking them up and some insight as to why they vanished was
obtained using Regge's 6j symmetries relations.  In that method a
given 6j symbol was shown to be equal to one which had at least one
small number up to two.  For such 6j expressions there are analytical
expressions which can be used.

The wavefunctions of
the I=$\frac{13}{2}^-$ state in $^{43}$Ti ($^{43}$Sc) can be written as 
\begin{eqnarray}
\psi_1 =a [J_{\pi}=4, j_{\nu}=\frac{7}{2}]^{I=\frac{13}{2}}+b[J_{\pi}=6, j_{\nu}=\frac{7}{2}]^{I=\frac{13}{2}}  \\
\psi_2 =-b[J_{\pi}=4, j_{\nu}=\frac{7}{2}]^{I=\frac{13}{2}} +a[J_{\pi}=6, j_{\nu}=\frac{7}{2}]^{I=\frac{13}{2}}
\end{eqnarray}
$(a^2+b^2=1)$

In general a and b are finite.  However when the T=0 two body
interaction matrix elements are set equal to zero we find a=1 b=0
i.e. the eigenfunctions are $[J_{\pi}=4, j_{\nu}=\frac{7}{2}]^{I=\frac{13}{2}}$
and $[J_{\pi}=6, j_{\nu}=\frac{7}{2}]^{I=\frac{13}{2}}$.  Furthermore the first
$\frac{13}{2}^-$ state is degenerate with the I=$\frac{1}{2}^-$ state
$[J_{\pi}=4, j_{\nu}=\frac{7}{2}]^{I=\frac{1}{2}}$.  The fact that both the
$\frac{1}{2}^-$ and $\frac{13}{2}^-$ states have the same structure
$[4,\frac{7}{2}]$ gives us some insight into why they are degenerate.

There are two conditions to be met

a) no mixing condition
$
\left\{ \begin{array}{ccc}
 j     & j      & 4\\
 j     & \frac{13}{2}      & 6             \\
\end{array} \right\} 
=0$

b) degeneracy

$
\left\{ \begin{array}{ccc}
 j     & j      & 4\\
 j     & \frac{13}{2}      & 4             \\
\end{array} \right\} 
=
\left\{ \begin{array}{ccc}
 j     & j      & 4\\
 j     & \frac{1}{2}      & 4             \\
\end{array} \right\} 
$
and
$
\left\{ \begin{array}{ccc}
 j     & j      & 6\\
 j     & \frac{13}{2}      & 6             \\
\end{array} \right\} 
=$
$
\left\{ \begin{array}{ccc}
 j     & j      & 6\\
 j     & \frac{17}{2}      & 6             \\
\end{array} \right\} 
=$
$
\left\{ \begin{array}{ccc}
 j     & j      & 6\\
 j     & \frac{19}{2}      & 6             \\
\end{array} \right\} 
$

These conditions are true.  They can be generalized to

a) 
$
\left\{ \begin{array}{ccc}
 j     & j      & (2j-3)\\
 j     & (3j-4) & (2j-1)\\
\end{array} \right\} 
=0$\\
b)
$
\left\{ \begin{array}{ccc}
 j     & j      & (2j-1)\\
 j     & I & (2j-1)\\
\end{array} \right\}=
\frac{(-1)^{2j}}{(8j-2)}$ for I = (3j-1), (3j-2), and (3j-4).
These hold for both half integer and integer j.

From degeneracy of the $9_1^+$ and $10^+_1$ states in $^{44}$Ti we get
conditions on the 9j symbols $ \left\{ \begin{array}{ccc} j & j & 6\\
j & j & 6\\ 4 & 6 & I \\ \end{array} \right\}=0$ for I=10

This decouples states with $(J_{\pi},J_{\nu})$ =(4,6)+(6,4) from states (6,6)

This can be generalized to
$
\left\{ \begin{array}{ccc}
 j     & j      & (2j-1)\\
 j     & j &  (2j-1)\\
 (2j-1)    & (2j-3)         &    (4j-4)   \\
 \end{array} \right\}=0$

We also get the diagonal condition 
$ \left\{ \begin{array}{ccc} 
j & j &(2j-3)\\ 
j & j & (2j-1)\\ 
(2j-3)& (2j-1)& I \\ 
\end{array}\right\}=
\frac{1}{16(2j-2)(2j-1)-12}
=\frac{1}{4(2j-5)(4j-1)}$ for I =
(4j-4),(4j-5), and (4j-7).  (10, 9 and 7 in the above example).

Note that we are dealing only with T=$\frac{1}{2}$ states in $^{43}$Ti
and T=0 states in $^{44}$Ti.  In the next section we will show how
these conditions arise.  One clue is that the degenerate states in
$^{43}$Ti occur for angular momentum for which there are no (single j)
T=$\frac{3}{2}$ states and in $^{44}$Ti for angular momentum for which
there are no T=2 states.  Note that when the T=0 two body matrix
elements are set equal to zero we can classify the states by the dual
quantum number $(J_{\pi},J_{\nu})$.

\section{Talmi's method of cfp to Pauli forbidden states}

In order to explain the properties of 6j and 9j symbols in the
previous section we shall take what may appear a strange detour and
consider systems of \underline{identical} particles.  Although we are concerned
with degeneracies of T=$\frac{1}{2}$ states and T=0 states we shall now examine
T=$\frac{3}{2}$ states in $^{43}$Ca and T=2 states in $^{44}$Ca.

The Pauli principle imposes severe constraints on the $j^3$
configuration of 3 neutrons.  Although 3 non-identical particles could
couple to a maximum spin of $\frac{21}{2}$, for identical particles
the maximum is $\frac{15}{2}$. This is easy to see by trying to construct the
maximum of $j_3$.  To satisfy the Pauli principle
$M_{max}=\frac{7}{2}+\frac{5}{2}+\frac{3}{2}=\frac{15}{2}$.  Thus we can form a state I=$\frac{15}{2}$ M=$\frac{15}{2}$.  One
then notes there is only one way to form M=$\frac{13}{2}$ namely $\frac{7}{2}+\frac{5}{2}+\frac{1}{2}$

This must correspond to I=$\frac{15}{2}$ M=$\frac{13}{2}$ state.  So there is no M=$\frac{13}{2}$
left for the I=$\frac{13}{2}$ state. Hence the I=$\frac{13}{2}$ state is forbidden by the Pauli
principle.

This is all in Talmi's book ~\cite{talmi}.  Of particular interest is
the fact stated therein by Talmi that one can extract useful information by
trying to construct coefficients of fractional parentage for this
forbidden I=$\frac{13}{2}$ state and then setting the cfp equal to
zero.

We can form a I=$\frac{13}{2}$ state as follows:
\begin{equation}
\psi^{0}=[[j(1)j(2)]I_0j(3)]I
\end{equation}
here $I_0$ must be even so that particles 1 and 2 are antisymmetrized.
For I=$\frac{13}{2}$, $I_0$ can be either 4 or 6.  Note that this
wavefunction is not overall antisymmetric.  The antisymmetric wave
function is $N(1-P_{13}-P_{23})\psi^{0}$.  The act of
antisymmetrization means that $I_0$ is merely a starting index and is
usually not unique.  The cfp expansion is
$\psi_{antisymmetric}=\Sigma_{I_1}(j^2I_1j|\}j^{3}I)[[j(1) j(2)
]^{I_1}j(3)]I$.  Each term in the expansion is not antisymmetric but
the total expression is.  Clearly the act of putting particle 3 to the
extreme right is going to involve some Racah coefficients.

In fact the explicit expression is 
\begin{equation}
(j^{2}I_1j|\}j^{3}I)=N [\delta_{I_1I_0} + 2 \sqrt{2I_0+1)(2I_1+1)} 
\left\{ \begin{array}{ccc}
 j     & j      & I_0\\
 I     & j      & I_1\\
\end{array} \right\}]
\end{equation}
\begin{equation}
N=[3+6(2I_0+1)
\left\{ \begin{array}{ccc}
 j     & j      & I_0\\
 I     & j      & I_0             \\
\end{array} \right\}]^{-1/2}
\end{equation}
However as Talmi points our since the I=$\frac{13}{2}$ state does not exist all the above cfp must vanish.~\cite{talmi}

Let us pick the starting $I_0$ to be 4, then the $I_1$ can be either 4 or 6. If
$I_1$ equals 4 we get 
$
\left\{ \begin{array}{ccc}
 j     & j      & 4\\
 I     & j      & 4             \\
\end{array} \right\}
=\frac{-1}{18}$ for I=$\frac{13}{2}$.  Here we note that this 6j is
independent of I for all I that are forbidden by the Pauli principle
i.e. $\frac{1}{2}$ and $\frac{13}{2}$. If $I_1$ equals 6 we get $ \left\{
\begin{array}{ccc} j & j & 4\\ I & j & 6 \\
\end{array} \right\}
=0$ But these are exactly the conditions we derived for the
degeneracies of I=$\frac{1}{2}^-$ and $\frac{13}{2}^-$ (T=$\frac{1}{2}$)
states of $^{43}$Sc.  The vanishing of the second 6j gives the
decoupling of $[J_{\pi}=4,j_{\nu}=\frac{7}{2}]^{I=\frac{13}{2}}$ from $[J_{\pi}=6,j_{\nu}=\frac{7}{2}
]^{I=\frac{13}{2}}$ and the first 6j gives the condition that $[
J_{\pi}=4,j_{\nu}=\frac{7}{2}]^{I=\frac{13}{2}}$ state and
$[J_{\pi}=4,j_{\nu}=\frac{7}{2}]^{I=\frac{1}{2}}$ are degenerate.

These results can be easily generalized to the following relations: 
$
\left\{ \begin{array}{ccc}
 j     & j      & (2j-3)\\
 (3j-4)     & j      & (2j-1)             \\
\end{array} \right\}
=0$
this holds for both half integer and integer j.  By a Regge symmetry~\cite{regge} this 6j is also equal to 
$
\left\{ \begin{array}{ccc}
 (2j-2)     & (2j-3)      & 2\\
 (2j-2)     & (2j-1)      & (2j-2) \\
\end{array} \right\}
$.  There are simple analytic expressions for 6j symbols where one of
the entries is two or less so it is easy to show that this 6j vanishes.
The other condition is
$
\left\{ \begin{array}{ccc}
 j     & j      & I_0\\
 I     & j      & I_0\\
\end{array} \right\}
=\frac{(-1)^{2j}}{2(2I_0+1)}$ for states I which are forbidden by the Pauli
principle for three identical particles.  The important thing to point
out is that for such states the 6j is independent of the total angular
momentum I.  This is necessary for states of different I to be
degenerate.

For states of angular momentum I which are not allowed by the Pauli principle
there are alternative ways of writing some of the relations 
i.e. 
$
\left\{ \begin{array}{ccc}
 j     & j      & (2j-1)\\
 j     & I      & (2j-1)\\
\end{array} \right\}
=\frac{(-1)^{2j}}{(8j-2)}$
for I=(3j-1), (3j-2), and (3j-4).

The argument extends to integer j.  Consider as an example j=4.  Then the vanishing 6j $
\left\{ \begin{array}{ccc}
 j     & j      & (2j-3)\\
 (3j-4)     & j      & (2j-1)\\
\end{array} \right\}$ = $
\left\{ \begin{array}{ccc}
 4     & 4      & 5\\
 8     & 4      & 7\\
\end{array} \right\}$ 
where we can regard j now as the orbital angular momentum.  We try to
form an antisymmetric state for I=8 using
(1-P$_{13}$-P$_{23}$)[[44]$^5$4]8.  But we can show that there is no
L=8 antisymmetric state of 3 L=4 particles.  The highest M one can
have is M=4+3+2=9.  We can thus have an L=9 M=9 state.  There is only
one way to form M=8 ie 4+3+1.  This must correspond to the L=9 M=8
states.  This leaves no M=8 to form an L=8 state.  Hence the cfp for
such a state much vanish.  Thus for the starting angular momentum
L$_A=5$ we must have the cfp for L$_1$=7 to vanish i.e.  $
\left\{ \begin{array}{ccc}
 4     & 4      & 5\\
 8     & 4      & 7\\
\end{array} \right\}=0$.

\section{Extension of the arguments to two particle fractional parentage coefficients}

To explain the degeneracies of certain T=0 states in $^{44}$Ti, we
consider a system of 4 identical particles i.e. $^{44}$Ca and as an
obvious extension of the previous section we consider the two particle
cfp's.

They are defined     
by 
\begin{equation}
\Psi_{antisymmetrized}=\Sigma_{I_1,I_2}(j^{n-2}I_1j^{2}I_{2}|\}
j^{n}I)[(j^{n-2})I_1(j^{2})I_2]^{I}
\end{equation} 

We can form an antisymmetric state
for 4 neutrons in the single j shell as follows 
\begin{equation}
(1-P_{13}-P_{23}-P_{14}-P_{24})[[j(1)j(2)]I_A[j(3)j(4)]I_B]^{I}
\end{equation} 
with the starting angular momentum $I_A$ and $I_B$ even.
 
Consider the
$P_{13}$ term - call it C
\begin{equation}
C=-[[j(3)j(2)]I_A[j(1)j(4)]I_B]^{I}
\end{equation} 
We must bring particle 3 to the right side and
particle 1 to the left side.
Thus 
\begin{equation} 
C=-(-1)^{1+I_A}
[[j(2)j(3)]I_A[j(1)j(4)]I_B]^{I}
\end{equation}
\begin{equation}
C= \Sigma_{I_1,I_2}(-1)^{I_A+I_1+1} <(jj)I_A(jj)I_B|(jj)I_1(jj)I_2>^{I}
  [[j(1)j(2)]I_1[j(3)j(4)]I_2]^{I}
\end{equation}
At this point $I_1$ and 
$I_2$ can be even or odd.  However when we perform
the complete antisymmetrization $I_1$ and $I_2$ will be even.
We find 
\begin{equation}
(j^2I_1j^2I_2|\}j^4I)=N[\delta_{I_AI_1}\delta_{I_BI_2}-4\sqrt{(2I_A+1)(2I_B+1)((2I_1+1)(2I_2+1)}
\left\{ \begin{array}{ccc}
 j     & j      & I_A\\
 j     & j      & I_B             \\
  I_1          & I_2       &   I            \\
\end{array} \right\}]
\end{equation}

Let us consider the states with angular momentum I=10. We recall that
when all the T=0 two body matrix elements were set equal to zero there
was a decoupling of the states for which ($J_{\pi}$,$J_{\nu}$) were
(6,4) and (4,6) from the state (6,6).  This demanded that $\left\{
\begin{array}{ccc} j & j & 6\\ j & j & 6 \\ 4 & 6 & 10 \\
\end{array} \right\}=0$
as indeed it is.

The only allowed states for 4 neutrons in the $f_{\frac{7}{2}}$ shell
are\\ 
v=0 I =0\\ 
v=2 I=2,4,6\\ 
v=4 I=2,4,5,8\\ 
In the above v is the seniority quantum number.
Since I =10
is not allowed all the 2 particle cfp above must vanish.

Let us choose the starting angular momentum to be $I_A=4$ and $I_B=6$.
We get two conditions from the fact that the two particle cfp vanish

a)$I_1=4$ $I_2 =6$ 
$ \left\{ \begin{array}{ccc}
 j     & j      & 4\\
 j     & j      & 6            \\
 4  & 6       &   10            \\
\end{array} \right\}=\frac{1}{468}$

b) $I_1=6$ $I_2=6$
$
\left\{ \begin{array}{ccc}
 j     & j      & 4\\
 j     & j      & 6             \\
 6        & 6       &   I            \\
\end{array} \right\}=0$

The second condition is the one required for the decoupling of
$(J_{\pi},J_{\nu})$=(4,6)+(6,4) from (6,6).  The first condition is better
stated as 
$\left\{ \begin{array}{ccc}
 j     & j      & 4\\
 j     & j      & 6             \\
 4     & 6      & I           \\
\end{array} \right\}$
is independent of I for states which are not allowed by the Pauli
principle for 4 neutrons and which can be formed by the vector sum
$\vec{4}+\vec{6}$.  These are 3, 7, 9, and 10, precisely the states
for which we get degeneracies.

The above results can be generalized to 
$ \left\{ \begin{array}{ccc}
 j     & j      & (2j-1)\\
 j     & j      & (2j-1)       \\
 (2j-1)          & (2j-3)      &  (4j-4)           \\
\end{array} \right\}=0$ 
and 
$
\left\{ \begin{array}{ccc}
 j     & j      & (2j-3)\\
 j     & j      & (2j-1)             \\
 (2j-3)         & (2j-1)       &   I            \\
\end{array} \right\}=\frac{1}{16(2j-2)(2j-1)-12}=\frac{1}{4(4j-5)(4j-1)}$.  
Both of these relations hold for integer j as well as half integer j.
We can now make a different choice for the starting angular momentum
i.e. $I_A=I_B=6$ and consider the case where $I_1=6$ and $I_2=6$.
This ``diagonal'' term is independent of I for the states not allowed
by the Pauli principle, a condition necessary for degeneracy.  Without
going into the detail, this will lead to the result that the second
10$^+$ state in $^{44}$Ti is degenerate with the unique 12$^{+}$
state.

\section{Electromagnetic Transitions between even spin and 
odd spin states in $^{44}$Ti}

We now consider the electromagnetic transitions from the even spin -
even parity states to the odd spin-even parity states.  We take the
example of the $10_1^+ \rightarrow 9^+_1$.  The transition can occur
either via the M1 or E2 modes.  In the single j shell approximation
B(M1) will vanish.  This is because the M1 operator in the single j
shell becomes
\begin{equation}
\sqrt{\frac{3}{4 \pi}} (\frac{g_{j \pi} + g_{j \nu}}{2}) \vec{I}
\end{equation}
and the total angular momentum $\vec{I}$ cannot induce M1 transitions.
Even when configuration mixing is included the value of B(M1) will be
very small because the isoscalar M1 coupling is very weak.  This is
due to the fact that the proton and neutron have magnetic moments of
opposite sign.  

The E2 transitions are not strongly collective but they should still
dominate over the M1's.  The values of B(E2)$_{10 \rightarrow 9}$ are 
7.167 e$^{2}$fm$^{4}$ in the single  shell and 6.589 e$^{2}$fm$^{4}$
in a full f-p calculation.  The corresponding values of B(M1) are zero and 
1.07x10$^{-4}$ $\mu_{n}^{2}$.

\section{Summary}

In summary we find that we can explain the many degeneracies in single
j shell calculations for T=$\frac{1}{2}$ states in
$^{43}$Ti($^{43}$Sc) and T=0 states in $^{44}$Ti when all the two body
interaction matrix elements are set equal to zero.  These degeneracies
involve states of angular momentum I, which because of the
Pauli principle cannot exist for T=$\frac{3}{2}$ in
$^{43}$Ti($^{43}$Sc) or T=2 states in $^{44}$Ti (or alternatively
for systems of identical particles e.g. $^{43}$Ca,
$^{44}$Ca).  These degeneracies require certain 6j and 9j symbols to
vanish and others to be equal.  These conditions can be derived using
Talmi's method of calculating coefficients of fractional parentage to
states that are forbidden by the Pauli principle and setting them
equal to zero.  One can apply this method to other single j shells
e.g. $g_{\frac{9}{2}}$.  Thus for $^{97}$Cd there would be degeneracy
of I=$\frac{19}{2}^+$, $\frac{23}{2}^+$, and $\frac{25}{2}^+$ and in
$^{96}$Cd I=13$^{+}$ and 14$^{+}$.

This work was supported by the U.S. Dept. of Energy under Grant
No. DE-FG02-95ER-40940 and one of us by GK-12 NSF9979491
(SJQR).

\end{document}